\documentclass{aastex631}

\begin{document}

\title{Size and Albedo Constraints for (152830) Dinkinesh Using WISE Data}

\correspondingauthor{Kiana D. McFadden}
\email{kmcfadden@arizona.edu}

\author[0000-0003-1383-2503]{Kiana D. McFadden}
\affiliation{Lunar and Planetary Laboratory, University of Arizona, Tucson, AZ, USA 
1629 E University Blvd 
Tucson, AZ 85721-0092 USA}

\author[0000-0002-7578-3885]{Amy K. Mainzer }
\affiliation{Lunar and Planetary Laboratory, University of Arizona, Tucson, AZ, USA 
1629 E University Blvd 
Tucson, AZ 85721-0092 USA}

\author[0000-0003-2638-720X]{Joseph R. Masiero}
\affiliation{Caltech/IPAC, 1200 E. California Blvd, MC 100-22, Pasadena, CA 91125 USA}

\author[0000-0001-9542-0953]{James M. Bauer}
\affiliation{Dept. of Astronomy, Univ. of Maryland, College Park, MD}

\author[0000-0002-0077-2305]{Roc M. Cutri}
\affiliation{Caltech/IPAC, 1200 E. California Blvd, MC 100-22, Pasadena, CA 91125 USA}

\author[0000-0003-1876-9988]{Dar Dahlen}
\affiliation{Caltech/IPAC, 1200 E. California Blvd, MC 100-22, Pasadena, CA 91125 USA}

\author[0000-0002-8532-9395]{Frank J. Masci}
\affiliation{Caltech/IPAC, 1200 E. California Blvd, MC 100-22, Pasadena, CA 91125 USA}

\author[0000-0002-5736-1857]{Jana Pittichov\'{a}}
\affiliation{Jet Propulsion Laboratory,
California Institute of Technology,
Pasadena, CA 91109, USA}

\author[0000-0001-5766-8819]{Akash Satpathy}
\affiliation{Lunar and Planetary Laboratory, University of Arizona, Tucson, AZ, USA 
1629 E University Blvd 
Tucson, AZ 85721-0092 USA}

\author[0000-0001-5058-1593]{Edward L. Wright}
\affiliation{University of California, Los Angeles, Los Angeles, CA, USA}



\begin{abstract}

Probing small main-belt asteroids provides insight into their formation and evolution through multiple dynamical and collisional processes. These asteroids also overlap in size with the potentially hazardous near-earth object population and supply the majority of these objects. The Lucy mission will provide an opportunity for study of a small main-belt asteroid, (152830) Dinkinesh. The spacecraft will perform a flyby of this object on November 1, 2023, in preparation for its mission to the Jupiter Trojan asteroids. We employed aperture photometry on stacked frames of Dinkinesh obtained by the Wide-field-Infrared Survey Explorer and performed thermal modeling on a detection at 12 $\mu$m to compute diameter and albedo values. Through this method, we determined Dinkinesh has an effective spherical diameter of $0.76^{+0.11}_{-0.21}$ km and a visual geometric albedo of $0.27^{+0.25}_{-0.06}$ at the 16th and 84th percentiles. This albedo is consistent with typical stony (S-type) asteroids. 

\end{abstract}



\section{Overview} \label{sec:intro} 

Our solar system formed from a molecular cloud of dust and gas, and as the solar nebula flattened into a disk and the protosun formed, dust grains began to condense \citep{Weidenschilling.1997a}. These grains eventually formed planetesimals through accretion; some planetesimals formed the planets we know today while others may have stopped growing at smaller sizes. The main-belt asteroids' compositions depend on where they were formed within the protoplanetary disk and how they were mixed in the early solar system \citep{Morbidelli.2009a,DeMeo.2015a}. Mixing may have occurred through planetary migration or through streaming instabilities \citep{Carrera.2015a,Tsiganis.2005a}. In addition, collisional cascades resulted in showers of smaller fragments that could migrate more rapidly due to non-gravitational forces \citep[i.e. Yarkovsky drift;][]{Bottke.2005a, Bottke.2005b, Gomes.2005a}. These smaller fragments can also re-accrete into rubble piles like Bennu, Ryugu, and Itokawa \citep[e.g.,][]{Nakamura.2022a}. The small main-belt asteroids we observe today in our solar system are likely the results from a combination of these processes \citep{Bottke.2015a}. By studying small main-belt asteroids we can gain insight into their formation and subsequent evolution.

Small main-belt asteroids are also important to study because they feed the current near-earth object (NEO) population, some of which have the potential to impact the Earth, and because their size scale overlaps with the largest NEOs \citep{Alvarez.1980a}. However, observational effects make it hard to study small main-belt asteroids because they are often too faint, far away, or small. On its way to explore the Jupiter Trojan asteroids, NASA's Lucy spacecraft will fly by the small main-belt asteroid (152830) Dinkinesh on November 1, 2023. The Lucy mission will collect observations using its thermal infrared spectrometer, high-resolution panchromatic imager, infrared imaging spectrometer, and color camera \citep{Levison.2021a}. These instruments will provide a diameter of the asteroid as well as a resolved shape model. Lucy will also obtain information on the visual geometric albedo, spin state, color maps, and constraints on volume and densities. Dinkinesh will become the smallest main-belt asteroid to have detailed fly-by data.

Dinkinesh was discovered by the Lincoln Near-Earth Asteroid Program \citep[LINEAR;][]{Stokes.2000a} survey in 1999. It has multi-epoch observations of sufficient quantity to provide a well constrained orbit, which makes Dinkinesh a suitable spacecraft fly-by target. \citet{Bolin.2023a} were able to obtain spectroscopic observations using the Keck and Gemini-South telescopes that indicated that Dinkinesh is a S-type/Sq-type asteroid, and using the albedo range from \citet{Mainzer.2011b} and the absolute visible magnitude of $H_{V} = 17.62\pm0.04$ mag from \citet{Mottola.2023a}, \citet{Bolin.2023a} calculated an effective diameter range of $0.67-0.96$ km based upon the relationship $D_{eff}= 1329 p_{V}^{-1/2} 10^{-H_{V}/5}$ km \citep{Fowler.1992a}. These results are in good agreement with \citet{deLeon.2023a}, who obtained $H_{V}=17.48\pm0.05$ mag and $0.16<G_{V}<0.23$ mag, giving them an effective spherical diameter range of $0.542-1.309$ km, and with \citet{Mottola.2023a} who found $G=0.378$ mag giving a diameter range of $0.66-1.36$ km at $2\sigma$. \citet{Mottola.2023a} also obtained light curve photometry of Dinkinesh and found a rotational period of $P=52.67\pm0.04$ hours, indicating a slowly rotating object.

In this paper, we report an independent measurement of size and visual geometric albedo of Dinkinesh that we derived from thermal modeling based on 12 $\mu$m stacked photometry from the Wide-field Infrared Survey Explorer taken in 2010 \citep{Wright.2010a,Mainzer.2011a}.

\section{Observations} \label{sec:obs}

The Wide-field Infrared Survey Explorer \citep[WISE;][]{Wright.2010a} is a NASA mission that created an all-sky map at four wavelengths spanning 3.4, 4.6, 12 and 22 $\mu$m (denoted W1, W2, W3, and W4, respectively). It launched in December 14, 2009, at which point all four bands were available. WISE completed its fully cryogenic primary mission on August 6, 2010. Although its primary scientific objective was not observing asteroids, it was able to detect $\sim$190,000 of them using automated detection software \citep{Mainzer.2011a}. Following the successful completion of its prime mission, the WISE spacecraft was reactivated in late 2013 and renamed the Near-Earth-Object Wide-field Infrared Survey Explorer \citep[NEOWISE;][]{Mainzer.2014a}  with its primary focus being characterization of NEOs using the two short-wavelength channels that were still operational, W1 and W2. Observations at thermal wavelengths allow for determination of asteroid diameters using thermal models, and if visible light data are available in addition to thermal measurements, it is possible to determine visible albedo as well as diameter.

\begin{figure}[ht!]
\plotone{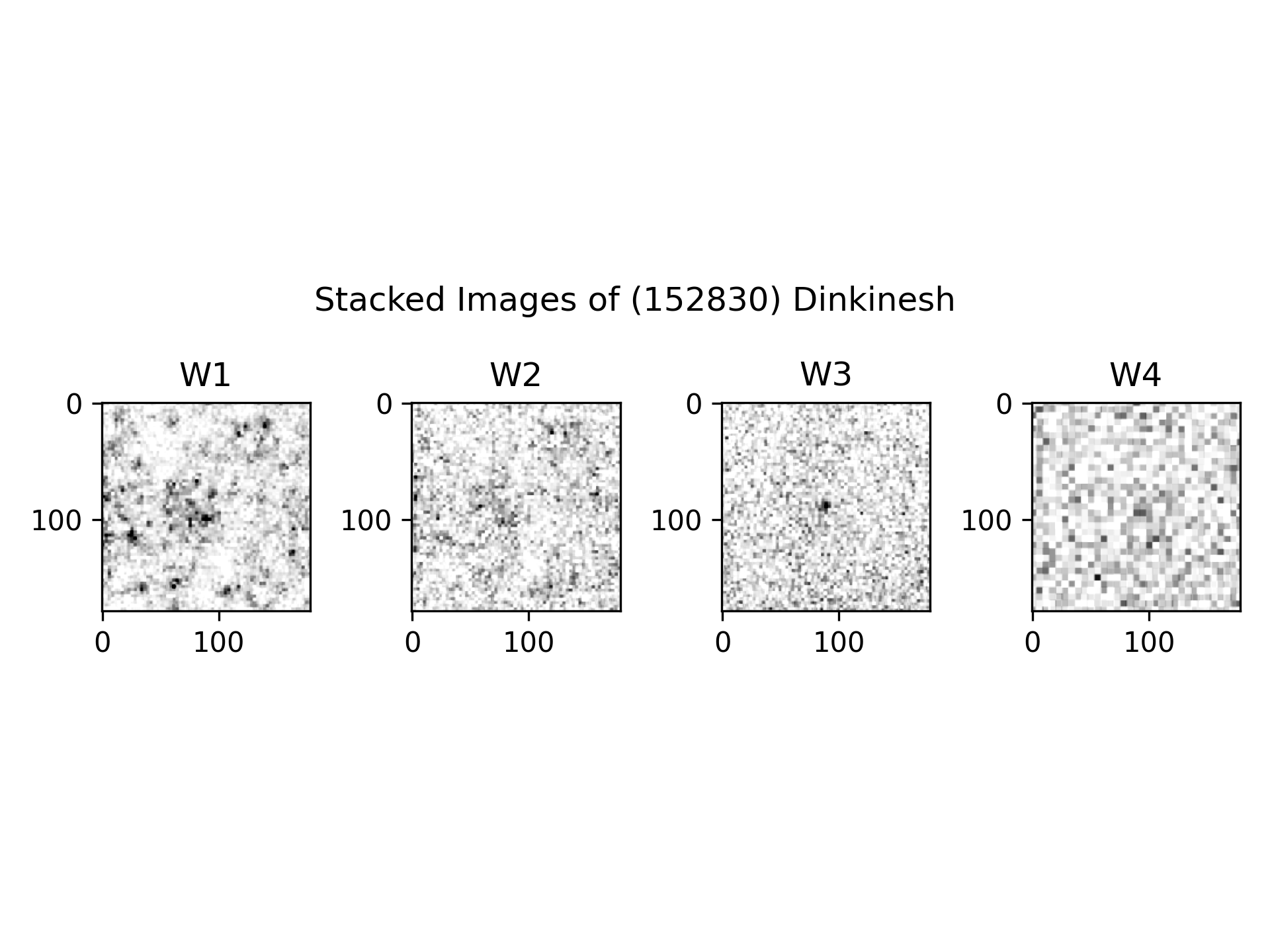}
\caption{Stacked image photometry of Dinkinesh in WISE bands W1, W2, W3, and W4.
\label{fig:dinkinesh_stacked_image}}
\end{figure}

During the primary fully cryogenic survey, WISE obtained 19 independent sets of exposures covering the position of Dinkinesh, 17 of which were of good quality. Table \ref{tab:dink_obs} summarizes our observations. The asteroid was generally too faint to be detected in the individual exposures, but we coadded the good quality frames by registering them to the position of the asteroid at the time of each exposure and coadding the shifted images. Exposures in which the predicted position of the asteroid coincided with a background star or galaxy were also eliminated prior to the coadd, based on examination of the AllWISE Source Catalog \citep{Cutri.2014a}. Exposures taken within 20$^{\circ}$ of the Moon were also excluded. The remaining 17 exposures were combined using the Image Co-addition with Optional Resolution Enhancement \citep[ICORE;][]{Masci.2009a, Masci.2013a} algorithm.\footnote{https://irsa.ipac.caltech.edu/applications/ICORE/docs/icore.pdf} The resulting coadded images in the four WISE bands are shown in Figure \ref{fig:dinkinesh_stacked_image}. The object is detected with a signal-to-noise ratio (SNR) of $\sim$5 at 12 $\mu$m. We searched NEOWISE W1 and W2 exposures taken after the fully cryogenic mission and did not find any detections.

\begin{deluxetable}{ccccccccc}
\label{tab:dink_obs}
\tablecaption{Stacked image photometry of Dinkinesh for aperture three. MJD = Modified Julian Date. $R_{helio}$ = radial heliocentric distance to the object. $R_{obs}$ = radial distance to the object from the observer.}
\setcounter{table}{0}
%\tablenum{1
\tablehead{\colhead{Minimum MJD} & \colhead{Median MJD} & \colhead{Maximum MJD} & \colhead{RA} & \colhead{Dec} & \colhead{ 12 $\mu$m Magnitude} & \colhead{SNR} & \colhead{$R_{helio}$} & \colhead{$R_{obs}$} \\ 
\colhead{Days} & \colhead{Days} & \colhead{Days} & \colhead{Deg} & \colhead{Deg} & \colhead{} & \colhead{} & \colhead{au } & \colhead{au}\\ }
\startdata
55274.2502 & 55275.0441 & 55275.7720 & 95.6132 & 25.8412 & 11.50 $\pm$ 0.23 &  4.6 &  1.98 &  1.62\\
\enddata
\end{deluxetable}

\section{Methods}\label{sec:metho}

We performed aperture photometry on the coadded images of Dinkinesh using a set of nested circular apertures in band W3 with radii of 11, 12, 14, 18, 22, 26, and 33 arcsec and a large annular region to estimate the level of the sky signal. We did not perform photometry on the W1 and W2 images because they were contaminated with stars due to low galactic latitude (Dinkinesh was 5$^{\circ}$ off the galactic plane at the time of observations). The object was too faint to produce a significant detection in the stacked W4 frames, although these data can be used to set an upper brightness limit.

\begin{figure}[ht!]
\plotone{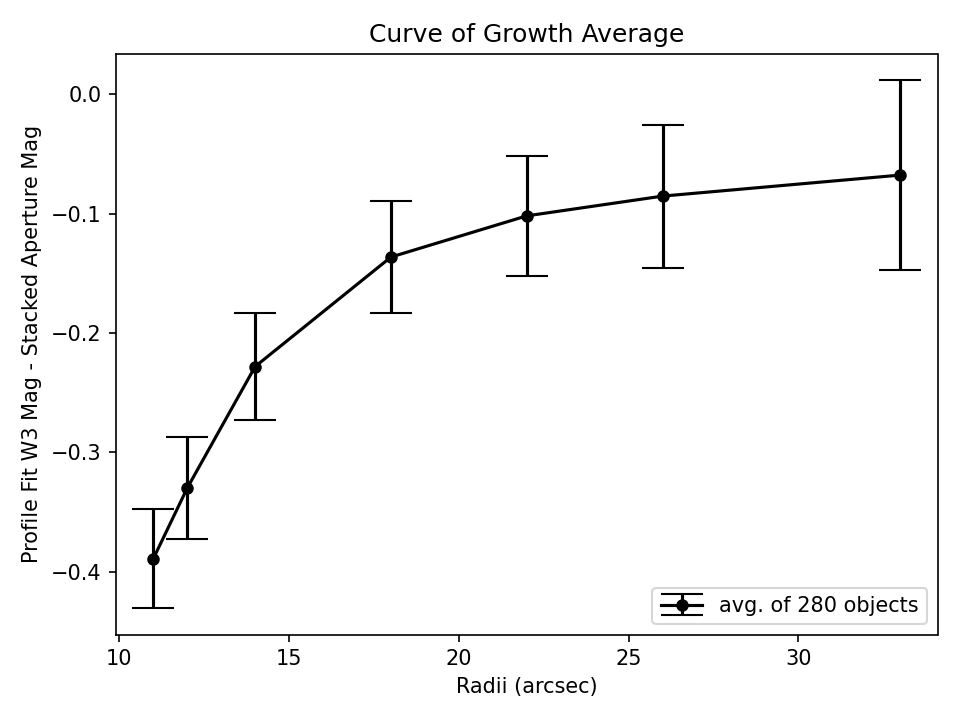}
\caption{We computed the average profile fit magnitude for 280 bright, unsaturated objects minus the stacked aperture photometry for each object. We then averaged over the 280 objects' offsets to compute curve of growth for band W3. The offset for aperture two was $-0.33$ mag, and the offset for aperture three was $-0.23$ mag.
\label{fig:dinkinesh_curve_of_growth}}
\end{figure}

We calibrated the aperture source photometry back to a well-established photometric standard using the profile-fit photometry from WISE. We made stacks of 280 bright and unsaturated asteroids (SNRs $> 30.0$ and W3 $>4.0$ magnitude) using the ICORE algorithm. Next, we computed aperture photometry for each object's stacked image and compared it to that object's profile-fit photometry. We employed an algorithm that flagged any extraneous sources such as stars, cosmic rays, diffraction spikes, optical artifacts, and smeared images from the profile-fit photometry. The rejection criteria included \emph{dtanneal} $>$ 2000 sec, \emph{cc$\_$flags} = 0, \emph{qi$\_$fact} $\neq 0$, \emph{rchi2} $<5$, and \emph{moon$\_$sep} $>$ 20 $^{\circ}$. Any detections from AllWISE that were within 5 arcsec of the asteroid's predicted position in a frame and within 1 mag of the asteroid's signal were excluded. For each of the 280 objects, we computed the offset between the profile fit W3 magnitude and all seven stacked aperture magnitudes. We then computed the average of the offsets for each aperture size to obtain the curve of growth, shown in Figure \ref{fig:dinkinesh_curve_of_growth}. We sought to minimize the background noise while also avoiding an overly small aperture size that would induce quantization effects; we selected aperture three with a 14 arcsec radius as the best balance. The limiting photometric calibration uncertainty is 0.03 mag for WISE \citep{Cutri.2012a}.\footnote{https://wise2.ipac.caltech.edu/docs/release/allsky/expsup/sec6\_3c.html} We derived the aperture correction by using enough bright asteroids to reduce the correction term's uncertainty ($0.23\pm0.05$ mag; Figure \ref{fig:dinkinesh_curve_of_growth}) to much less than the photometric measurement uncertainty from Dinkinesh, which was $\sim\pm$0.2 mag. We applied the aperture correction term to Dinkinesh’s aperture magnitude to correct back to a calibrated profile fit W3 magnitude (Table \ref{tab:dink_obs}). We tested apertures two and three (12 and 14 arcsec, respectively), and the results were the same within the statistical uncertainty.

\subsection{Thermal Modeling}

The Near-Earth Asteroid Thermal Model, NEATM, is a simple thermal model that improves upon earlier versions such as the standard thermal model and the fast rotating model \citep{Harris.1998a}. The standard thermal model approximates an asteroid as a non-rotating sphere while solar insolation and the surface temperatures are in instantaneous equilibrium \citep{Lebofsky.1986a}. The fast-rotating model \citep{Lebofsky.1989a} assumes that the temperature is distributed in uniform bands in latitude all around the asteroid. Our study uses NEATM because previous work \citep{Mainzer.2011b,Wright.2018a, Masiero.2021a} has demonstrated that the diameter measurements determined from NEOWISE using NEATM are generally reliable to within $\sim$10\%, and it is computationally efficient. NEATM makes a number of assumptions to model the temperature distribution across the asteroid's surface. The NEATM model assumes that the asteroid is not rotating and that the night-side has a temperature of 0 K. Our implementation of NEATM approximates a sphere by using a sufficient number of triangular facets arranged in a Fibonacci lattice.\footnote{https://extremelearning.com.au/how-to-evenly-distribute-points-on-a-sphere-more-effectively-than-the-canonical-fibonacci-lattice/} This model also incorporates a beaming parameter $\eta$ which is allowed to vary to account for additional thermal parameters such as thermal inertia and surface roughness, when multiple thermally dominated bands are available. We find the best fit solution using the Python \emph{basinhopper} {\citep{Wales.1997a,Virtanen.2020a} package and find the uncertainties using the Monte Carlo Markov Chain approach implemented by the Python \emph{emcee} package \citep{Foreman-Mackey.2013a}.

We apply NEATM to Dinkinesh to obtain its effective spherical diameter and albedo values. We set the emissivity value to 0.9 based on meteoritical studies \citep{Bates.2021a,Ostrowski.2020a}. We adopt $H_{V} = 17.62\pm0.04$ mag based on the photometric measurements of \citet{Mottola.2023a}. Similarly, we adopted $ G = 0.378$ mag from \citet{Mottola.2023a} but tested G over the full range 0.08 to 0.38 mag and found that this did not make a statistically significant difference in the results because diameter is only weakly dependent on G. Since we do not have multiple measurements at more than one thermally dominated band, we assumed a beaming value $\eta$ of 1.00 $\pm 0.1$ based on \citet{Masiero.2011a,Masiero.2014a}.

\section{Results/Discussion}\label{sec:disc}

We observed Dinkinesh over the course of 36.5 hours compared to its rotational period of 52.67 hours \citep{Mottola.2023a}. The long duration of our observations with WISE allows us to better constrain its effective spherical diameter by observing over most of its full rotational cycle. Through our thermal models, we determined that Dinkinesh has an effective spherical diameter of $0.76^{+0.11}_{-0.21}$ km and a visual geometric albedo of $0.27^{+0.25}_{-0.06}$ with the uncertainties specified at the 16th and 84th percentiles. Figure \ref{fig:dinkinesh_results} shows the best fit solution from the thermal models. The albedo is consistent with the range of typical values for S-type asteroids \citep{Mainzer.2011d}. 

We also used a spherical thermophysical model (TPM) as described in \citet{Wright.2007a} and \citet{Koren.2015a} to determine the 
diameter of Dinkinesh, which gave a smaller estimate of 621$^{+92}_{-117}$ m. Most of the diameter difference between the TPM and NEATM is caused by the unusually long rotation period ($P=52.67$ hours) of Dinkinesh, 
which is an order of magnitude longer than the median for typical objects with diameters near 700 m. This is smaller than our NEATM-fit size, but they are consistent within uncertainties. We prefer NEATM due to the limited data available.

\begin{figure}[ht!]
\plotone{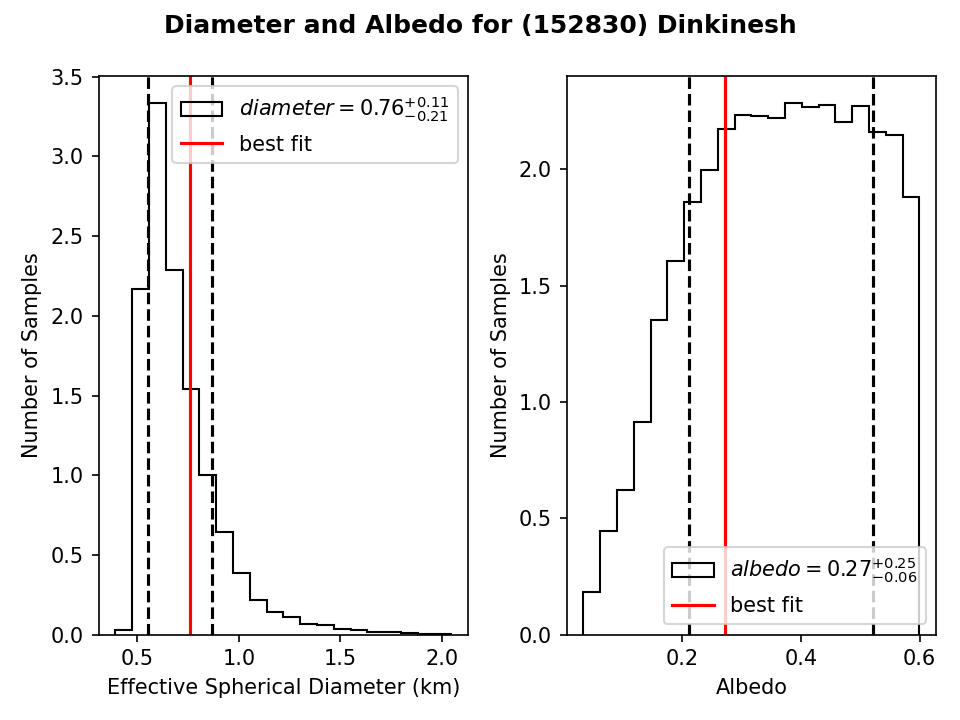}
\caption{We determined the effective spherical diameter (left) and visual geometric albedo (right) for Dinkinesh. The histograms represent the probability distribution functions from the Monte Carlo Markov Chain trials. The red line represents the best fit from the Monte Carlo Markov Chain trials to probe the probability distribution for both parameters. The two dashed lines represent the 16th and 84th percentiles.
\label{fig:dinkinesh_results}}
\end{figure}

\subsection{Potential Asteroid Family Association}
Asteroid family associations can be made by comparing orbital precession rates or proper orbital elements \citep{Hirayama.1918a,Milani.1994a,Knezevic.2002a}. Asteroid (152830) Dinkinesh is not formally part of any asteroid
family.\footnote{see AstDys:
\textit{https://newton.spacedys.com/astdys/index.php?pc=5}} However, the synthetic proper orbital elements \citep{Knezevic.2003a} for Dinkinesh show a very close match to those
for asteroid (8) Flora in proper semi-major axis (2.191 vs 2.201 au), proper eccentricity (0.1482 vs 0.1449), rate of perihelion precession \textit{g} (32.810 vs 32.017 arcsec/year), and rate of ascending node precession \textit{s} (-35.502 vs -35.511 arcsec/year), deviating only in proper inclination (1.600$^\circ$ vs 5.574$^\circ$).

This implies that there is a potential link between Dinkinesh and the Flora family, with the large deviation in inclination being a result of the initial velocity due to the family-forming impact, or due to the seasonal Yarkovsky effect \citep{Vokrouhlick.1999a} that has a large out-of-plane component for
objects with obliquities that are neither parallel nor orthogonal to their orbital plane \citep{Bottke.2002a}. This link is further supported by the consistency between the spectral taxonomy of Dinkinesh and Flora.

\section{Conclusions}\label{concl}

We obtained photometry at 12 $\mu$m of the small main-belt asteroid (152830) Dinkinesh by coadding multiple independent exposures obtained by WISE in March 2010. We used two thermal models, NEATM and TPM, to obtain diameter and visual geometric values. These values are in good agreement with the predicted diameter and albedo ranges from \citet{Bolin.2023a,deLeon.2023a}, and \citet{Mottola.2023a}. Based on the axial ratio of $a/b\sim1.43$ derived from the light curve amplitude by \citep{Mottola.2023a}, our spherical-equivalent size of $0.76^{+0.11}_{-0.21}$ km would correspond to axial sizes of $2a=0.96$ km and $2b=2c=0.67$ km, assuming that our measurements fully sample a half-rotation of Dinkinesh. From the WISE data, we determined that the visual geometric albedo is $0.27^{+0.25}_{-0.06}$, which is consistent with typical S-type asteroids. The data that will be obtained from Lucy's flyby of Dinkinesh will provide us with a detailed look at a very small main-belt asteroid. That, in turn, will help us to better understand the provenance of such objects.

\section{Acknowledgements}
\begin{acknowledgments}

This publication makes use of data products from the Wide-field Infrared Survey Explorer, which is a joint project of the University of California, Los Angeles, and the Jet Propulsion Laboratory/California Institute of Technology, funded by the National Aeronautics and Space Administration. This publication also makes use of data products from NEOWISE, which is a project of the Jet Propulsion Laboratory/California Institute of Technology, funded by the Planetary Science Division of the National Aeronautics and Space Administration.

This research has made use of the NASA/IPAC Infrared Science Archive, which is funded by the National Aeronautics and Space Administration and operated by the California Institute of Technology.

This research has made use of data and/or services provided by the International Astronomical Union's Minor Planet Center.

The work of J.P. was carried out at the Jet Propulsion Laboratory, California Institute of Technology, under a contract with the National Aeronautics and Space Administration.

Dataset usage:

\dataset[WISE All-Sky 4-band Single-Exposure Images]{https://www.ipac.caltech.edu/doi/irsa/10.26131/IRSA152}

\dataset[AllWISE Source Catalog]
{https://www.ipac.caltech.edu/doi/irsa/10.26131/IRSA1}

\dataset[NEOWISE-R Single Exposure (L1b) Source Table]
{https://www.ipac.caltech.edu/doi/irsa/10.26131/IRSA144}

\end{acknowledgments}


\bibliography{references.bib}{}
\bibliographystyle{aasjournal}



\end{document}